\documentclass[twoside,leqno,twocolumn]{article}

\usepackage{siamproceedings}

\usepackage{algorithmic}
\usepackage{amsfonts}
\usepackage{amsmath}
\usepackage{amssymb}
\usepackage{calc}
\usepackage{cite}
\usepackage{enumitem}
\usepackage{float}
\usepackage{forloop}
\usepackage{listings}
\usepackage{microtype}
\usepackage{pgf}
\usepackage{tikz}
\usepackage{wrapfig}
\usepackage{url}
\usepackage[bf]{caption}

\newcommand{\dd}{{\mathinner{.\,.}}}

\newcommand{\emptystring}{\epsilon}
\newcommand{\bwt}[1]{{\operatorname{BWT}_{#1}}}

\newcommand{\lcp}[1]{{\operatorname{LCP}_{#1}}}
\newcommand{\plcp}[1]{{\operatorname{PLCP}_{#1}}}

\newcommand{\sa}[1]{{\operatorname{SA}_{#1}}}

\newcommand{\dol}{\text{\normalfont $\texttt{\$}$}}

\newcommand{\rank}[3]{{\operatorname{rank}_{#1,#2}(#3)}}
\newcommand{\select}[3]{{\operatorname{select}_{#1,#2}(#3)}}

\newcommand{\popcount}[1]{{\operatorname{popcount}(#1)}}
\newcommand{\encodelength}[1]{{\operatorname{enclen}(#1)}}
\newcommand{\hblock}[2]{{\operatorname{hblock}_{#1}(#2)}}
\newcommand{\hblockrank}[3]{{\operatorname{hblockrank}_{#1,#2}(#3)}}
\newcommand{\hblockoffset}[2]{{\operatorname{hblockoffset}_{#1}(#2)}}
\newcommand{\sblock}[2]{{\operatorname{sblock}_{#1}(#2)}}
\newcommand{\sblockrank}[3]{{\operatorname{sblockrank}_{#1,#2}(#3)}}
\newcommand{\sblockoffset}[2]{{\operatorname{sblockoffset}_{#1}(#2)}}
\newcommand{\block}[2]{{\operatorname{block}_{#1}(#2)}}
\newcommand{\blockrank}[3]{{\operatorname{blockrank}_{#1,#2}(#3)}}
\newcommand{\blockoffset}[2]{{\operatorname{blockoffset}_{#1}(#2)}}
\newcommand{\blockones}[2]{{\operatorname{blockones}_{#1}(#2)}}
\newcommand{\blockencodelength}[2]{{\operatorname{blockenclen}_{#1}(#2)}}
\newcommand{\specialbit}[2]{{\operatorname{specialbit}_{#1}(#2)}}
\newcommand{\kparam}{{k_{\rm param}}}
\newcommand{\kinterval}{{k_{\rm interval}}}
\newcommand{\mmax}{{m_{\rm max}}}

\definecolor{mypink}{RGB}{199, 21 133}
\hypersetup{
  hidelinks,
  bookmarksopen,
  bookmarksnumbered,
  colorlinks,
  urlcolor = blue,
  linkcolor = mypink,
  citecolor = blue
}

\allowdisplaybreaks

\begin{document}

\title{Engineering Select Support for Hybrid Bitvectors
\thanks{Partially funded by the NSF CAREER Award 2337891.}}

\author{
  Eric Chiu\thanks{
    Department of Computer Science,
    Stony Brook University,
    Stony Brook, NY, USA,
    \texttt{echiu@cs.stonybrook.edu}.
  }
  \and
  Dominik Kempa\thanks{
    Department of Computer Science,
    Stony Brook University,
    Stony Brook, NY, USA,
    \texttt{kempa@cs.stonybrook.edu}.
  }
}

\date{\vspace{-0.5cm}}

\maketitle

\begin{abstract}
    One of the central problems in the design of compressed data
    structures is the efficient support for rank and select queries on
    bitvectors. These two operations form the backbone of more complex
    data structures (such as wavelet trees) used for the compact
    representation of texts, trees, graphs, or grids. Their efficient
    implementation is one of the most frequently studied problems in
    compressed data structures.

    One effective solution is the so-called hybrid bitvector
    implementation, which partitions the input bitvector into blocks
    and adaptively selects an encoding method—such as run-length,
    plain, or minority encoding—based on local redundancy. Experiments
    have shown that hybrid bitvectors achieve excellent all-around
    performance on repetitive and non-repetitive inputs.

    However, current implementations support only rank queries (i.e.,
    counting the number of ones up to a given position), and lack
    support for select queries. This limitation significantly
    restricts their applicability. In this paper, we propose a method
    to add support for select queries to hybrid bitvectors, and we
    conduct an extensive set of experiments. Our results show that
    hybrid bitvectors offer excellent performance—matching the speed
    of the fastest and the space efficiency of the most compact
    existing bitvectors.
\end{abstract}

\section{Introduction}\label{sec:intro}

Compressed data structures are essential for the efficient storage and
access of increasingly large datasets. In domains such as
bioinformatics\textemdash where data volumes are measured in
terabytes\textemdash classical data structures are too large to be
practical~\cite{1000genomesproject,100kgp}. As a result, compressed
data structures have been developed as space-efficient alternatives
that retain good query performance. As datasets continue to grow in
size, so will the demand for increasingly compact data structures that
support even faster queries.

Many compressed data structures, when traced down to their most basic
components, consist of bitvectors augmented with support for basic
operations~\cite{FerraginaM05,wavelettreehuff,sada_cst,FerraginaMMN04,BelazzouguiN14,KempaK23}.
\emph{Rank} and \emph{select} queries are among the most commonly used
basic operations~\cite{navarrobook}, and the space usage and query
performance of those operations often determine the space and speed of
the whole data structure.
To further improve space efficiency, these bitvectors can be
compressed. While a plain bitvector with constant-time rank and select
support requires $n + o(n)$ bits~\cite{munro1996tables}, a
\textit{compressed bitvector} can achieve the same in a fraction of
that space~\cite{rrrvector,patrascu2008succincter}.

Although there are many ways to compress a bitvector, no single
approach is optimal for all binary sequences. For example, the RRR
bitvector~\cite{rrrvector} compresses best when there is an uneven
distribution of zeros and ones throughout; the SD
bitvector~\cite{sdarray} compresses best when there are very few one
bits but struggles when the bitvector is dense; finally, a run-length
encoding, as in the RLWT~\cite{rlcsa_and_rlbwt,rlbwt_complete},
compresses best when the number of runs is low but is not
space-efficient otherwise. In practice, bitvectors often contain
distinct regions, each of which prefers a different
encoding~\cite{hybvector}.

Consequently, in~\cite{hybvector}, K\"{a}rkk\"{a}inen et al. proposed
the hybrid bitvector, a compressed bitvector that individually encodes
each block with the method that best reduces its space. They
demonstrated that it consistently produced one of the smallest
FM-indices with one of the fastest query times for both repetitive and
nonrepetitive texts. However, select queries on the hybrid bitvector
were not implemented, which has limited its usage.

In this paper, we present a method to enable select queries on the
hybrid bitvector, thereby achieving full bitvector functionality.  The
hybrid bitvector can now assist in compressed data structures that
require a bitvector with select support, whereas previously it could
not.  Finally, we show that the hybrid bitvector with select query
support retains its excellent all-round performance across repetitive
and nonrepetitive texts.

\paragraph{Related Work}

Rank and select queries on bitvectors have been studied extensively.
Uncompressed bitvectors can support constant-time rank and select
queries in $o(n)$ additional bits~\cite{munro1996tables}.  Efficient
implementations are provided by Navarro and
Providel~\cite{navarro2012fast}, Zhou et al.~\cite{zhou2013space}, and
Gog and Petri~\cite{gog2014optimized}.

Compressed bitvectors can also support constant-time rank and select
queries in $o(n)$ additional bits.  The RRR bitvector~\cite{rrrvector}
is one such example.  It achieves implicit $k^{\text{th}}$ order
entropy compression~\cite{implicit_compression_boosting} by
individually compressing each block into a variable-length pair.  The
SD bitvector~\cite{sdarray} is another such compressed bitvector.  It
applies an Elias--Fano encoding to the positions of ones in the
bitvector~\cite{elias, fano}.

Our implementation extends SDSL's \texttt{hyb\_vector}
class~\cite{sdsl}, which was adapted from the hybrid bitvector
implementation by K\"{a}rkk\"{a}inen et al.~\cite{hybvector}. The
hybrid bitvector has been studied for indices that do not require
select queries, such
as~\cite{hybvector,hybvectorinfmindex,fasterminuter,hybindex}.

\section{Preliminaries}

A string is a sequence of characters from an alphabet.  We use
$\Sigma$ to denote an alphabet, $\sigma$ to denote the size of
$\Sigma$, and $S \in \Sigma^n$ to denote a string of length $n$ from
the alphabet $\Sigma$.  A bitvector is a string from the binary
alphabet $\{0, 1\}$.  Given strings $S_1$ and $S_2$, $S_1S_2$ denotes
the concatenation of $S_1$ and $S_2$.

Given $S \in \Sigma^n$ and $i,j\in[1 \dd n]$, $S[i]$ denotes the
$i\textsuperscript{th}$ character in $S$, $S[i \dd j]$ denotes the
substring $S[i] \cdots S[j]$, $S[1 \dd i]$ denotes a prefix of $S$,
and $S[i \dd n]$ denotes a suffix of $S$. The empty string is denoted
$\emptystring$. We will often assume that $S[|S|]$ is a special sentinel
character (denoted $\dol$), which is the smallest character in $\Sigma$.

We use $\preceq$ to denote a total order on the alphabet $\Sigma$
extended to the so-called \emph{lexicographic} order on $\Sigma^{*}$
so that $S_1 \preceq S_2$ holds for any $S_1 \in \Sigma^{n_1}$ and $S_2 \in \Sigma^{n_2}$
if either $S_1$ is a prefix of $S_2$, or it holds
$S_1[1 \dd i) = S_2[1 \dd i)$ and $S_1[i] \prec S_2[i]$ for some
$i \in [1 \dd \min(n_1, n_2)]$.

\subsection{Suffix Array (SA)}

The suffix array~\cite{suffixarray} of $S \in \Sigma^n$, denoted
$\sa{S}$, is a permutation of $[1 \dd n]$ such that $\sa{S}[i]$ is the
starting index of the $i$\textsuperscript{th} lexicographically
smallest suffix of $S$, i.e., such that $S[\sa{S}[i] \dd n] \prec
S[\sa{S}[i+1] \dd n]$ holds for every $i \in [1 \dd n)$;
see \cref{suffixarrayexample} for an example.

\begin{figure}[t!]
  \centering
  \begin{tabular}{rrrrl}
    $i$ & $\sa{S}[i]$ & $\lcp{S}[i]$ & $\bwt{S}[i]$ & $S[\sa{S}[i] \dd n]$ \\
    \hline
    1 & 7 & 0 & \texttt{a}  & \texttt{\$} \\
    2 & 6 & 0 & \texttt{n}  & \texttt{a\$} \\
    3 & 4 & 1 & \texttt{n}  & \texttt{ana\$} \\
    4 & 2 & 3 & \texttt{b}  & \texttt{anana\$} \\
    5 & 1 & 0 & \texttt{\$} & \texttt{banana\$} \\
    6 & 5 & 0 & \texttt{a}  & \texttt{na\$} \\
    7 & 3 & 2 & \texttt{a}  & \texttt{nana\$}
  \end{tabular}
  \caption{The suffix array, LCP array, and BWT of string $S = \texttt{banana\$}$.}
  \label{suffixarrayexample}
\end{figure}

\subsection{Longest Common Prefix (LCP) Array}

The LCP array of $S \in \Sigma^n$, denoted $\lcp{S}$, is an array of
$n$ numbers where $\lcp{S}[i]$ is the length of the longest common
prefix between $S[\sa{S}[i] \dd n]$ and $S[\sa{S}[i-1] \dd n]$ for $i
\in [2 \dd n]$.  The edge case is $\lcp{S}[1] = 0$.  Together,
$\sa{S}$ and $\lcp{S}$ can solve complex pattern matching queries on
$S$~\cite{suffixarray}.  \Cref{suffixarrayexample} provides an example
of an LCP array.

\subsection{Burrows--Wheeler Transform (BWT)}

The BWT~\cite{burrows1994block} of $S \in \Sigma^n$, denoted
$\bwt{S}$, is a permutation of symbols in $S$ such that, for every $i \in [1 \dd n]$,
\[
  \bwt{S}[i] =
    \begin{cases}
      S[n] & \text{if } \sa{S}[i] = 1, \\
      S[\sa{S}[i]-1] & \text{otherwise.} \\
    \end{cases}
\]
The BWT a core component of the FM-index~\cite{fmindex}.
\Cref{suffixarrayexample} provides an example of a BWT.

\subsection{Rank and Select Queries}

Rank and select on $S \in \Sigma^n$ for character $c \in \Sigma$ is
defined as follows:
\begin{align*}
  \rank{S}{c}{i} &= \text{the number of $c$ in $S[1..i]$} \\
    &= |\{x \in [1 \dd i] : S[x] = c\}| \quad \forall i \in [1 \dd n] \\
  \select{S}{c}{j} &= \text{the position of the $j$\textsuperscript{th} instance of $c$ in $S$} \\
    &= \min\{x \in [1 \dd n] : \rank{S}{c}{x} = j\} \quad \\
    &\qquad \forall j \in [1 \dd \rank{S}{c}{n}]
\end{align*}
Bitvectors can support rank and select queries in constant time and
$o(n)$ additional space~\cite{munro1996tables}.

\subsection{Wavelet Tree}

A wavelet tree~\cite{wavelettree} of $S \in \Sigma^n$ is a binary tree
where each leaf corresponds to a symbol in $\Sigma$ and each internal
node represents a subsequence of $S$ with only the symbols for leaves
below that node.  Additionally, each internal node has a bitvector
with a zero/one if the symbol in the represented subsequence is in the
left/right subtree, respectively.

\begin{figure}[t!]
\centering
\begin{tikzpicture}
    \draw[draw=none] (0,0.5) rectangle ++(0.3,0.5) node[pos=0.5] {\strut\texttt{a}};
    \draw[draw=none] (0.3,0.5) rectangle ++(0.3,0.5) node[pos=0.5] {\strut\texttt{b}};
    \draw[draw=none] (0.6,0.5) rectangle ++(0.3,0.5) node[pos=0.5] {\strut\texttt{c}};
    \draw[draw=none] (0.9,0.5) rectangle ++(0.3,0.5) node[pos=0.5] {\strut\texttt{a}};
    \draw[draw=none] (1.2,0.5) rectangle ++(0.3,0.5) node[pos=0.5] {\strut\texttt{b}};
    \draw[draw=none] (1.5,0.5) rectangle ++(0.3,0.5) node[pos=0.5] {\strut\texttt{a}};
    \draw[draw=none] (1.8,0.5) rectangle ++(0.3,0.5) node[pos=0.5] {\strut\texttt{c}};
    \draw (0,0) rectangle ++(0.3,0.5) node[pos=0.5] {0};
    \draw (0.3,0) rectangle ++(0.3,0.5) node[pos=0.5] {1};
    \draw (0.6,0) rectangle ++(0.3,0.5) node[pos=0.5] {1};
    \draw (0.9,0) rectangle ++(0.3,0.5) node[pos=0.5] {0};
    \draw (1.2,0) rectangle ++(0.3,0.5) node[pos=0.5] {1};
    \draw (1.5,0) rectangle ++(0.3,0.5) node[pos=0.5] {0};
    \draw (1.8,0) rectangle ++(0.3,0.5) node[pos=0.5] {1};

    \node[circle,draw] (a) at (0.3,-0.75) {\texttt{a}};

    \draw[draw=none] (1.05,-1.0) rectangle ++(0.3,0.5) node[pos=0.5] {\strut\texttt{b}};
    \draw[draw=none] (1.35,-1.0) rectangle ++(0.3,0.5) node[pos=0.5] {\strut\texttt{c}};
    \draw[draw=none] (1.65,-1.0) rectangle ++(0.3,0.5) node[pos=0.5] {\strut\texttt{b}};
    \draw[draw=none] (1.95,-1.0) rectangle ++(0.3,0.5) node[pos=0.5] {\strut\texttt{c}};
    \draw (1.05,-1.5) rectangle ++(0.3,0.5) node[pos=0.5] {0};
    \draw (1.35,-1.5) rectangle ++(0.3,0.5) node[pos=0.5] {1};
    \draw[fill=none] (1.65,-1.5) rectangle ++(0.3,0.5) node[pos=0.5] {0};
    \draw (1.95,-1.5) rectangle ++(0.3,0.5) node[pos=0.5] {1};
    \node[circle,draw] (b) at (1.2,-2.25) {\texttt{b}};
    \node[circle,draw] (c) at (2.1,-2.25) {\texttt{c}};

    \draw (1.05,0) -- (a);
    \draw (1.05,0) -- (1.65,-0.5);
    \draw (1.65,-1.5) -- (b);
    \draw (1.65,-1.5) -- (c);

\end{tikzpicture}
\caption{Wavelet tree for $S = \texttt{abcabac}$.}
\label{wavelettreeexample}
\end{figure}
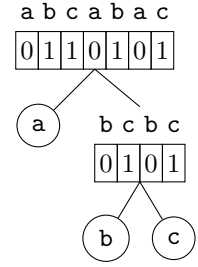

A wavelet tree reduces rank and select queries on general strings to
rank and select queries on bitvectors.  \Cref{wavelettreeexample}
provides a visualization of a wavelet tree.

\section{Hybrid Bitvector}

In this section, we describe the existing hybrid bitvector implementation, \texttt{hyb\_vector}, from SDSL~\cite{hybvector,sdsl}. 
We introduce the notation to later describe how we added select query support.

\subsection{Basic Structure}

\begin{definition}
    The $i^\text{th}$ $b$-bit block in bitvector $B \in \{0,1\}^n$ is the substring $B((i-1)b,\min\{ib, n\}]$.
\end{definition}

To compress bitvector $B \in \{0,1\}^n$, 
the hybrid bitvector partitions $B$ into $b$-bit blocks 
and individually compresses each block with the most space-efficient option from the following selection of encoding methods:

\begin{enumerate}
    \item Minority encoding: a list of minority bit positions in the block. 
    \item Run-length encoding: a list of run endings in the block, where a run ending is the index of the last bit of the run.
    \item Plain encoding: the uncompressed block. 
\end{enumerate}

Blocks are organized into $b_sb$-bit blocks, called superblocks,
and furthermore, into $b_hb$-bit blocks, called hyperblocks. 
The block size is $b = 2^8$, 
the number of blocks per superblock is $b_s \in \{8, 16, 32, 64\}$, 
and the number of blocks per hyperblock is $b_h = 2^{23}$.

The hybrid bitvector has three components:  
\begin{enumerate}[itemsep=0pt]
    \item an array of hyperblock headers $A_H$,
    \item an array of superblock and block headers $A_S$, and 
    \item an array of variable-sized block encodings $A_E$.
\end{enumerate}

\begin{definition}
    Let $\encodelength{B}$ denote the total length of block encodings for bitvector $B$.
\end{definition}

\begin{definition} We define utility functions for hyperblocks, superblocks, and blocks:
\begin{itemize}
    \item For the $i$\textsuperscript{th} hyperblock:
    \begin{align*}
        \hblock{B}{i} &= B(b_hb(i-1) \dd b_hbi] \\
        \hblockrank{B}{c}{i} &= \rank{B}{c}{b_hb(i-1)} \\
        \hblockoffset{B}{i} &= \encodelength{B[1 \dd b_hb(i-1)]}
    \end{align*}
    \item For the $i$\textsuperscript{th} superblock:
    \begin{align*}
        \sblock{B}{i} &= B(b_sb(i-1) \dd b_sbi] \\
        \sblockrank{B}{c}{i} &= \rank{B}{c}{b_sb(i-1)} \\
        &\quad - \hblockrank{B}{c}{\lfloor\tfrac{b_s(i-1)}{b_h}\rfloor + 1} \\
        \sblockoffset{B}{i} &= \encodelength{B[1 \dd b_sb(i-1)]} \\
        &\quad - \hblockoffset{B}{\lfloor\tfrac{b_s(i-1)}{b_h}\rfloor + 1}
    \end{align*}
    \item For the $i$\textsuperscript{th} block:
    \begin{align*}
        \block{B}{i} &= B(b(i-1) \dd bi] \\
        \blockrank{B}{c}{i} &= \rank{B}{c}{b(i-1)} \\
        &\quad - \hblockrank{B}{c}{\lfloor\tfrac{i-1}{b_h}\rfloor + 1} \\ 
        &\quad - \sblockrank{B}{c}{\lfloor\tfrac{i-1}{b_s}\rfloor + 1} \\
        \blockoffset{B}{i} &= \encodelength{B[1 \dd b(i-1)]} \\
        &\quad - \hblockoffset{B}{\lfloor\tfrac{i-1}{b_h}\rfloor + 1} \\
        &\quad - \sblockoffset{B}{\lfloor\tfrac{i-1}{b_s}\rfloor + 1} \\
        \blockones{B}{i} &= \popcount{\block{B}{i}} \\
        \blockencodelength{B}{i} &= \encodelength{\block{B}{i}}
    \end{align*}
\end{itemize}
\end{definition}

The $i$\textsuperscript{th} hyperblock header gives context for hyperblock $\hblock{B}{i}$ and contains 
(1) $\hblockrank{B}{1}{i}$, the number of ones preceding this hyperblock, 
and 
(2) $\hblockoffset{B}{i}$, the length of all block encodings for blocks preceding this hyperblock. 

The array $A_S$ contains superblock headers and block headers such that each superblock header is followed by the $b_s$ block headers for blocks inside that superblock.

A superblock header is similar to the hyperblock header, but its values are local to the hyperblock. 
The $i$\textsuperscript{th} superblock header gives context for superblock $\sblock{B}{i}$ and contains 
(1) $\sblockrank{B}{1}{i}$, the number of ones preceding this superblock, but inside the same hyperblock,
(2) $\sblockoffset{B}{i}$, the length of all block encodings for blocks preceding this superblock, but inside the same hyperblock, and
(3) a flag indicating whether the superblock is uniform.

The $i$\textsuperscript{th} block header describes block $\block{B}{i}$ and contains 
(1) $\blockones{B}{i}$, the number of ones in this block, 
(2) $\blockencodelength{B}{i}$, the size of the encoding for this block, and 
(3) $\specialbit{B}{i}$, a special bit that gives additional information about the encoding.

Though $\blockrank{B}{1}{i}$ and $\blockoffset{B}{i}$ are not stored, 
they can be computed as the sum of $\blockones{B}{\cdot}$ and $\blockencodelength{B}{\cdot}$
from blocks preceding this block, but inside the same superblock. 

The $\specialbit{B}{i}$ is interpreted differently for each encoding.
For the minority encoding, $\specialbit{B}{i}$ identifies the minority bit.
For the run-length encoding, it identifies the first bit of that block.
For the plain encoding, it is ignored.

The encoding type of the $i$\textsuperscript{th} block can be identified by examining $\blockencodelength{B}{i}$.
If it equals $\min\{\blockones{B}{i}, b - \blockones{B}{i}\}$, then the block is minority encoded.
If it equals $\frac{b}{8}$, then the block is plain encoded.
Otherwise, the block is run-length encoded.

Block encodings are concatenated to form array $A_E$.
The encoding for the $i$\textsuperscript{th} block is
located at offset $\hblockoffset{B}{\lfloor \frac{i - 1}{b_h} \rfloor + 1} + \sblockoffset{B}{\lfloor \frac{i - 1}{b_s} \rfloor + 1} + \blockones{B}{i}$
with length $\blockencodelength{B}{i}$.

\subsection{Run-Length Encoding Optimization}

The hybrid bitvector's run-length encoding 
does not list the last two run endings, 
because the headers in $A_H$ and $A_S$ provide enough information.
We mention this optimization because it is relevant to our implementation
of hybrid bitvector select queries.

Here, we show that we can recover the full run-length encoding
even with the above optimization.
Suppose that the $i$\textsuperscript{th} block in $B \in \{0,1\}^n$ is run-length encoded 
with all but the last two run endings $(r_1,r_2,\ldots,r_{m-2})$.
That is, there are $m = \blockencodelength{B}{i} + 2$ runs in $\block{B}{i}$.
We know $r_m = b$. 
The set of run endings for runs containing 1-bits is
$D = \{x \in [1 \dd m] : 2 \mid (x+\specialbit{B}{i})\}$. We can compute $r_{m-1}$ by solving the following equation
(using $r_0 = 0$): $$\sum_{d\in D}(r_{d} - r_{d-1}) = \blockones{B}{i}$$
Hence, we are able to recover the full run-length encoding $(r_1,r_2,\ldots,r_{m})$
from only $(r_1,r_2,\ldots,r_{m-2})$ and block headers.

\section{Implementation of Select Queries on a Hybrid Bitvector}

In this section, we describe how we implemented select queries on the hybrid bitvector.
Our goal is to compute $\select{B}{c}{i}$ for bitvector $B \in \{0, 1\}^n$ and target bit $c \in \{0, 1\}$.

\subsection{Definitions}

To better describe lookup table $L_s$ in the next subsection, we define the following values:
\begin{itemize}[parsep=0.5ex,itemsep=0.5ex]
\item Let $\kparam$ denote a tunable parameter that restricts 
the space used by $L_s$ to at most $\lfloor \frac{n}{\kparam} \rfloor$ bits. 
We find $\kparam = 128$ works well in practice.
\item Let $w = 64$ denote the width of a superblock index in bits.
\item Let $\mmax = \max\{\lfloor\frac{n}{\kparam w}\rfloor, 2\}$ denote the maximum number of elements allowed in $L_s$.
\item Let $\kinterval = \lceil\frac{\rank{B}{c}{n}}{\mmax - 1}\rceil$ denote the gap between sampled $c$-bits in $L_s$.
\item Let $m = \lceil\frac{\rank{B}{c}{n}}{\kinterval}\rceil + 1$ denote the number of elements in $L_s$.
\end{itemize}

\subsection{Components}

To support select queries on the hybrid bitvector, we augment it with the following two components:
\begin{enumerate}[parsep=0.5ex,itemsep=0.5ex]
    \item sampling interval $\kinterval$ and
    \item lookup table $L_s$, which is an array of $m$ values such that
    $$L_s[i] = \begin{cases}
        \lceil \frac{n}{b_sb} \rceil & \text{if } i = m \\
        \lfloor\frac{\select{B}{c}{\kinterval(i-1) + 1}-1}{b_sb}\rfloor + 1 & \text{otherwise} 
    \end{cases}$$
    More simply, the $i$\textsuperscript{th} value in $L_s$ is the index of the superblock 
    that contains the $(\kinterval(i-1) + 1)$\textsuperscript{th} $c$-bit in $B$.
    Note that $L_s$, which has $m$ $w$-bit values, uses at most $\lfloor\frac{n}{\kparam}\rfloor = o(n)$ bits:
    \begin{align*}
        wm 
        &= w \left(\left\lceil\frac{\rank{B}{c}{n}}{\kinterval}\right\rceil + 1 \right) \\
        &= w \left(\left\lceil\frac{\rank{B}{c}{n}}{\lceil\rank{B}{c}{n}/(\mmax - 1)\rceil}\right\rceil + 1 \right) \\
        &\leq w \left(\left\lceil\frac{\rank{B}{c}{n}}{\rank{B}{c}{n}/(\mmax - 1)}\right\rceil + 1 \right) \\
        &= w\mmax
        = w \left\lfloor\frac{n}{\kparam w}\right\rfloor
        \leq \left\lfloor\frac{n}{\kparam}\right\rfloor
    \end{align*}
\end{enumerate}

\subsection{Answering Queries}

To compute $\select{B}{c}{q}$:
\begin{enumerate}
    \item First, we find the superblock index $i_s$ that contains the $q$\textsuperscript{th} $c$-bit in $B$.
    To do this, we compute:
    \begin{align*}
        u &= \left\lfloor\frac{q - 1}{\kinterval}\right\rfloor + 1 \\
        i_s' &= L_s[u] \\
        i_s'' &= L_s[u + 1]
    \end{align*}
    The lookup table and query are designed such that $i_s' \leq i_s \leq i_s''$.
    We locate $i_s \in [i_s' \dd i_s'']$ by binary search, where
    $i_s = \max\{x \in [i_s' \dd i_s''] : 
        \hblockrank{B}{c}{\lfloor\frac{b_s(x - 1)}{b_h}\rfloor + 1} 
        + \sblockrank{B}{c}{x} < q \}$
    From the superblock header, we extract $\sblockrank{B}{c}{i_s}$ and $\sblockoffset{B}{i_s}$.

    \item Next, we compute the hyperblock index $i_h$ that contains the $q$\textsuperscript{th} $c$-bit in $B$.
    The formula is $i_h = \lfloor\frac{b_s(i_s - 1)}{b_h}\rfloor$.
    From the hyperblock header, we extract $\hblockrank{B}{c}{i_h}$ and $\hblockoffset{B}{i_h}$.

    \item Then, we iterate over the $b_s$ block headers that follow the $i_s$\textsuperscript{th} superblock header in $A_S$ to find the block index $i_b$ that contains the $q$\textsuperscript{th} $c$-bit in $B$.
    We start with $i_b = b_s(i_s-1)$ and increment until $\hblockrank{B}{c}{i_h} + \sblockrank{B}{c}{i_s} + \blockrank{B}{c}{i_b} \geq q$. 
    Note that the block header does not provide $\blockrank{B}{1}{i_b}$, so we compute it as the sum of $\blockones{B}{\cdot}$ from all preceding blocks in the same superblock. 
    We also compute $\blockoffset{B}{i_b}$ as the sum of $\blockencodelength{B}{\cdot}$ from all preceding blocks in the superblock. 
    We keep $\blockrank{B}{1}{i_b}$, $\blockoffset{B}{i_b}$, $\blockencodelength{B}{i_b}$, and $\specialbit{B}{i_b}$ for the next step.  

    \item Let $B' = \block{B}{i_b}$. 
    We now have enough information to access the encoding of $B'$ and perform a localized select query on it. 
    Let $l = \blockencodelength{B}{i_b}$. 
    The block encoding, denoted by $R = (r_1,r_2,\ldots,r_l)$, is found at position $\hblockoffset{B}{i_h} + \sblockoffset{B}{i_s} + \blockoffset{B}{i_b}$ in $A_E$. 
    We determine the type of the encoding and compute $\select{B'}{c}{q'}$ where $q' = q - \hblockrank{B}{c}{i_h} - \sblockrank{B}{c}{i_s} - \blockrank{B}{c}{i_b}$.
    We describe how this is done for each type of encoding:

    \begin{enumerate}
        \item If the block is minority encoded, we use \cref{minorityencoding} to compute $\select{B'}{c}{q'}$.
        \item If the block is plain encoded,
        we use the built-in 64-bit popcount instructions and SDSL's 64-bit select function to compute $\select{B'}{c}{q'}$.
        \item If the block is run-length encoded, 
        we use \cref{runlengthencoding} to compute $\select{B'}{c}{q'}$.
    \end{enumerate}

    \item Finally, we return our result $(i_b-1)b + \select{B'}{c}{q'}$.
\end{enumerate}

\begin{algorithm}[t!]
\caption{Select Query on a Minority Encoding}
\label{minorityencoding}
\textbf{Input}: $c$, the target bit  \\
\hspace*{3.2em} $q'$, the local query index  \\
\hspace*{3.2em} $(r_1,r_2,\ldots,r_l)$, minority bit positions for $B'$ \\
\hspace*{3.2em} $\specialbit{B}{i_b}$, the minority bit \\
\textbf{Output}: $\select{B'}{c}{q'}$
\begin{algorithmic}[1]
\IF{$c = \specialbit{B}{i_b}$}
    \RETURN $r_{q'}$
\ELSE
    \STATE{Define $x := 0$ \qquad /* number of (1-c)-bits */}
    \WHILE{$x < l$ and $r_{x + 1} - x < q'$}
        \STATE{$x := x + 1$}
    \ENDWHILE
    \RETURN $x + q'$
\ENDIF
\end{algorithmic}
\end{algorithm}

\begin{algorithm}[t!]
\caption{Select Query on a Run-Length Encoding}
\label{runlengthencoding}
\textbf{Assumes}: $l > 0$. See \cref{implementation:optimization} if $l = 0$. \\
\textbf{Input}: $c$, target bit \\
\hspace*{3.2em} $q'$, local query index \\
\hspace*{3.2em} $(r_1,r_2,\ldots,r_l)$, all but the last two run endings \\
\hspace*{3.2em} $\specialbit{B}{i_b}$, the first bit in $B'$ \\
\hspace*{3.2em} $\blockones{B}{i_b}$, the number of ones in $B'$ \\
\textbf{Output}: $\select{B'}{c}{q'}$
\begin{algorithmic}[1]
\STATE{Define $a := 0$ \qquad /* result */}
\STATE{Define $u := q'$ \qquad /* remaining $c$ bits */}
\STATE{Define $x := 0$ \qquad /* run index minus one */}
\STATE{/* Start at a run of $(1-c)$ bits */}
\IF{$c = \specialbit{B}{i_b}$}
    \STATE{$a := a + \min\{r_1, u\}$}
    \STATE{$u := u - \min\{r_1, u\}$}
    \STATE{$x := x + 1$}
\ENDIF
\STATE{/* Iterate pairwise */}
\WHILE{$x + 1 < l$ and $u > 0$}
    \STATE{$a := a + r_{x+1} + \min\{r_{x+2} - r_{x+1}, u\}$}
    \STATE{$u := u - \min\{r_{x+2} - r_{x+1}, u\}$}
    \STATE{$x := x + 2$}
\ENDWHILE
\STATE{/* If the result is in the last two runs */}
\IF{$u > 0$}
    \IF{$x = l$}
        \STATE{$a := c (b - \blockones{B}{i_b})$}
        \STATE{$a := a + (1 - c) \blockones{B}{i_b}$}
        \STATE{$a := a + q'$}
    \ELSE
        \STATE{$a := r_l + u$}
    \ENDIF
\ENDIF
\RETURN{$a$}
\end{algorithmic}
\end{algorithm}

\subsection{Additional Optimizations}\label{implementation:optimization}

There are some additional optimizations not mentioned above.
In Step 1, we also check the flag that indicates a uniform superblock. 
If the flag is set, we can immediately return $b_sb(i_s-1) + (q - \hblockrank{B}{c}{i_h} - \sblockrank{B}{c}{i_s})$. 
Since $A_E$ is not accessed, we avoid a cache miss. 

In Step 4, $B'$ is run-length encoded and $\blockencodelength{B}{i_b} = 0$, then $B'$ has at most two runs.
If $c = \specialbit{B}{i_b}$ then the first run is a run of $c$-bits and $\select{B'}{c}{q'} = q'$. 
Otherwise, the second run contains the $c$-bits and the length of the first run must be included.
If $\specialbit{B}{i_b} = 1$, $\select{B'}{c}{q'} = \blockones{B}{i_b} + q'$.
Otherwise, $\select{B'}{c}{q'} = b - \blockones{B}{i_b} + q'$
Again, $A_E$ is not accessed, so we avoid a cache miss.

\section{Experimental Results}

In this section, we report the performance of the hybrid bitvector with select query support.\footnote{Our C++ implementation is available at \url{https://github.com/echiu12/hyb_vector_select/}.}
We consider two applications: PLCP queries and BWT select queries.

\subsection{Setup}

Our experiments were run on a 5.00GHz Intel(R) Core(TM) i7-1355U CPU with a 12MiB L3 cache, 
32GB system memory, and a 64-bit Linux Ubuntu 22.04.5 OS (kernel version 6.8.0-60-generic).

Our code is compiled with g++ version 11.4.0 
and flags \texttt{-Wall -static -DNDEBUG -msse4.2 -std=c++11 -funroll-loops -O3}. 
We measure time with \texttt{std::chrono::high\_resolution\_timer} function. 
We measure the size of data structures using a serialization function.

Our input texts are from the Pizza\&Chili website; see \cref{pizza&chili}. 
They are the same ten texts used in the original hybrid bitvector paper~\cite{hybvector}. 
These consist of five nonrepetitive texts and five repetitive texts.

\begin{table*}[t!]
\centering
\caption{
    Texts from Pizza\&Chili.
    The first five texts are 200MiB prefixes of nonrepetitive texts that can be downloaded from the standard corpus (\url{https://pizzachili.dcc.uchile.cl/texts.html}). 
    The rest are repetitive texts that can be downloaded from the repetitive corpus (\url{https://pizzachili.dcc.uchile.cl/repcorpus.html}). 
    Texts are sorted by the average length of runs in the BWT, $n/r$, which is a measure of repetitiveness.
}
\begin{tabular}{lrrrl}
\hline
Name & $\sigma$ & $n / 2^{20}$ & $n/r$ & Description \\
\hline
dna & 16 & 200 & 1.63 & DNA from the Human Genome Proj. \\
proteins & 25 & 200 & 1.93 & Proteins from the Swissprot database \\
english & 225 & 200 & 2.91 & English texts from Project Gutenberg \\
sources & 230 & 200 & 4.40 & Source files for Linux and GCC \\
dblp.xml & 96 & 200 & 7.09 & XML of the DBLP bibliography \\
\hline
para & 5 & 410 & 27 & 36 \textit{S. Paradoxus} DNA sequences \\
cere & 5 & 440 & 40 & 37 \textit{S. Cerevisae} DNA sequences \\
influenza & 15 & 148 & 51 & 78,041 Influenza DNA sequences \\
world\_leaders & 89 & 44 & 82 & World leaders directory by the CIA\\
kernel & 160 & 246 & 93 & 36 versions of Linux kernel sources\\
\hline
\end{tabular}
\label{pizza&chili}
\end{table*}

\subsection{Experiment 1: PLCP Queries}

The PLCP of string $S \in \Sigma^n$, 
denoted $\plcp{S}$, 
is a permutation of $\lcp{S}$ such that $\plcp{S}[j] = \lcp{S}[\sa{S}^{-1}[j]]$ where $j \in [1 \dd n]$. 
The succinct representation of $\plcp{S}$ is used to solve queries on $\lcp{S}$ for compressed suffix trees~\cite{plcp2n,sada_cst}.
In these applications, the $\plcp{S}$ is encoded as a dense bitvector $B_\plcp{S}$ of length $2n$ such that $\plcp{S}[j] = \select{B_\plcp{S}}{1}{j} - 2j$.
Typically, an uncompressed bitvector represents $B_\plcp{S}$, but we investigate alternatives.

In our first experiment, we compare the performance of PLCP queries using the following four bitvector types:
\begin{itemize}
    \item HYB: The hybrid bitvector with our implementation of select queries as described previously. 
    We consider blocks per superblock $b_s \in \{8, 16, 32, 64\}$.
    \item RRR: The RRR bitvector~\cite{rrrvector,gog2014optimized}, which zero-order compresses each block and decodes on the fly. 
    We consider block sizes 15, 31, 63, and 127. The implementation for block size 15 is highly optimized. 
    \item SD: The SD bitvector~\cite{sdarray}, which encodes the positions of ones with Elias--Fano encoding~\cite{elias, fano}.
    \item BV: The uncompressed bitvector.
\end{itemize}
Besides HYB, all bitvectors are available from SDSL~\cite{sdsl}.

When the above aliases are prefixed with ``PLCP\_'', they refer to the corresponding PLCP encoded with that bitvector type.
(e.g. PLCP\_HYB refers to the PLCP that is encoded with the hybrid bitvector).
For each text, we prepared $10^5$ indices from 1 to $n$, which were used to compute the average PLCP query time for each bitvector. 
We also measured the size of the PLCP bitvector relative to the text size and observed the space-time trade-off.

\begin{figure*}[t!]
\centering
\textbf{PLCP Query Time and Space Trade-offs}
\flushleft
\foreach \i in {1,...,4} {
    \includegraphics[page=\i,width=0.45\textwidth,trim={0 0.6cm 0 0},clip]{rsrc/bench_plcp__20250714_184300.pdf}
    \includegraphics[page=\numexpr\i+5\relax,width=0.45\textwidth,trim={0 0.6cm 0 0},clip]{rsrc/bench_plcp__20250714_184300.pdf}
}
\includegraphics[page=5,width=0.45\textwidth]{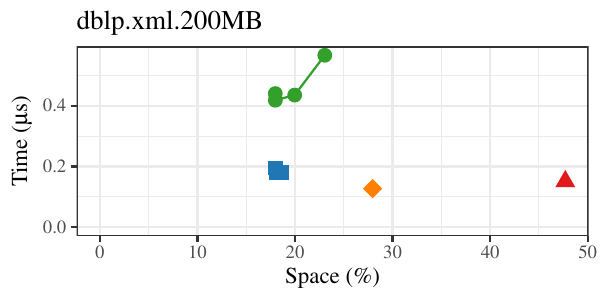}
\includegraphics[page=10,width=0.45\textwidth]{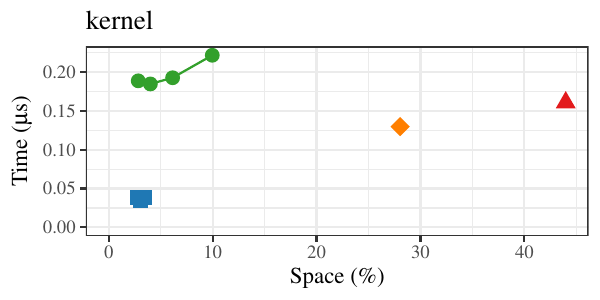}
\centering
\includegraphics[page=11,width=0.5\textwidth,trim={0, 1.5cm, 0, 1.5cm},clip]{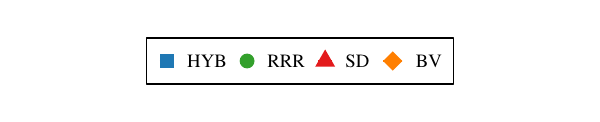}
\caption{
    Performance of PLCP queries for various bitvector types on nonrepetitive (left column) and repetitive (right column) texts.
    The query time is averaged over $10^5$ queries, and the space is relative to the text size. 
    PLCP\_HYB is comparable in size to PLCP\_RRR, while also being 2 to 3 times faster for all texts.
    Overall, PLCP\_HYB has excellent space-time trade-offs for all texts.
}
\label{plcp}
\end{figure*}

\Cref{plcp} shows the results of this experiment.
For both nonrepetitive and repetitive texts, PLCP\_HYB is consistently one of the smallest and one of the fastest. 
PLCP\_HYB is comparable in size to PLCP\_RRR while being 2 times faster for nonrepetitive texts and 3 times faster for repetitive texts.
In some cases, PLCP\_HYB is more space efficient than PLCP\_RRR.

PLCP\_HYB is smaller than PLCP\_BV by 5\% and 20\% of the text size for nonrepetitive and repetitive texts, respectively.
Between PLCP\_HYB and PLCP\_SD, these figures increase to 20\% and 35\%, respectively.
PLCP\_HYB is slower than PLCP\_BV for nonrepetitive texts\textemdash though still 2 times faster than PLCP\_RRR\textemdash 
and up to 2 times faster for repetitive texts. 
It achieves space similar to PLCP\_RRR without sacrificing as much in query speed.

Overall, PLCP\_HYB has superior space-time trade-offs for both repetitive and nonrepetitive texts.

\subsection{Experiment 2: BWT Select Queries}

Recall that the Burrows--Wheeler Transform~\cite{burrows1994block} of $S \in \Sigma^n$, 
denoted $\bwt{S}$, 
is a permutation of $S$ and serves as the core component of the FM-Index~\cite{fmindex}.
To support queries on $\bwt{S}$, it is represented as a wavelet tree,
which reduces select queries on $\bwt{S}$ 
to rank and select queries on bitvectors.

In our second experiment, we measure the performance of BWT select queries using different types of bitvectors.
We used the same bitvectors as in the first experiment, but we also considered two types of wavelet trees:
\begin{enumerate}
    \item HUFF: a Huffman-shaped wavelet tree~\cite{wavelettreehuff}.
    \item BLCD: a balanced wavelet tree~\cite{wavelettree}.
\end{enumerate}
Both wavelet trees are available from SDSL~\cite{sdsl}.

\begin{figure*}[t!]
\centering
\textbf{BWT Select Query Time and Space Trade-offs}
\flushleft
\foreach \i in {1,...,4} {
    \includegraphics[page=\i,width=0.45\textwidth,trim={0 0.6cm 0 0},clip]{rsrc/bench_bwt_select__20250714_184300.pdf}
    \includegraphics[page=\numexpr\i+5\relax,width=0.45\textwidth,trim={0 0.6cm 0 0},clip]{rsrc/bench_bwt_select__20250714_184300.pdf}
}
\includegraphics[page=5,width=0.45\textwidth]{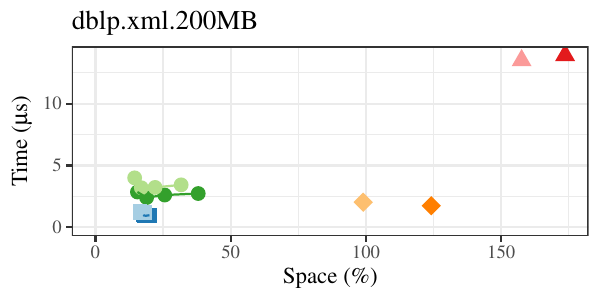}
\includegraphics[page=10,width=0.45\textwidth]{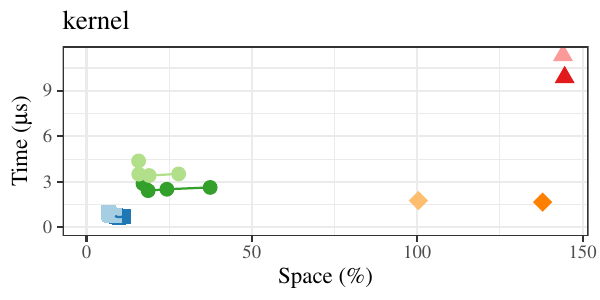}
\centering
\includegraphics[page=11,width=0.5\textwidth,trim={0, 1.5cm, 0, 1.5cm},clip]{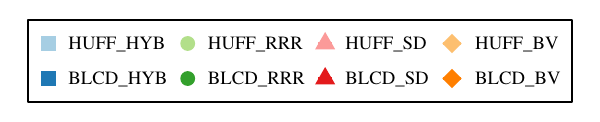}
\caption{
    Performance of BWT select queries for various wavelet tree and bitvector types 
    on nonrepetitive (left column) and repetitive (right column) texts.
    The query time is averaged over $10^5$ queries, and the space is relative to the text size. 
    HYB is comparable in size to RRR, while also being 2 times faster for all texts.
    Overall, HYB has excellent space-time trade-offs for all texts.
}
\label{bwt}
\end{figure*}

We refer to each wavelet tree and bitvector combination by their aliases separated with an underscore 
(e.g., HUFF\_HYB is a Huffman-shaped wavelet tree encoded in a hybrid bitvector). 
We refer to both Huffman and balanced wavelet trees for a particular bitvector using the bitvector's alias 
(e.g., HYB refers to both HUFF\_HYB and BLCD\_HYB).

\cref{bwt} shows the results of this experiment. 
HYB is slightly larger than RRR for nonrepetitive texts and smaller than RRR for repetitive texts, however, it is consistently 2 times faster.
Interestingly, for two repetitive texts in particular\textemdash para and cere\textemdash 
HYB outperforms RRR in both time and space by a large margin.

HYB is comparable to BV in query time, while also being consistently smaller than BV. 
The difference in size between HYB and BV ranges from 10\% to 125\% of the text size, increasing as the text becomes more repetitive. 
SD performs very poorly because select queries on both zero and one bits are required.

Similar to the previous experiment, HYB has excellent space-time trade-offs for both nonrepetitive and repetitive texts.
Its size is comparable to RRR and its speed is comparable to BV.

\bibliographystyle{siamplain}
\bibliography{refs}

\end{document}